\documentclass[runningheads]{llncs}
\usepackage[T1]{fontenc}
\usepackage{makecell}

\usepackage{graphicx,verbatim}
\usepackage{subcaption}
\usepackage{booktabs}
\usepackage{multirow}
 \usepackage{amssymb}
 \usepackage{amsmath}
\usepackage{threeparttable}
\usepackage{hyperref}
\usepackage{color,marvosym}

\begin{document}
\title{Probability-Invariant Random Walk Learning on Gyral Folding-Based Cortical Similarity Networks for Alzheimer's and Lewy Body Dementia Diagnosis}
\titlerunning{Probability-Invariant Random Walk Learning for Dementia Classification}
%

\author{Anonymized Authors}  
\authorrunning{Anonymized Author et al.}
\institute{Anonymized Affiliations \\
    \email{email@anonymized.com}}

\author{Minheng Chen \inst{1} \and Tong Chen \inst{1} \and Chao Cao \inst{1} \and Jing Zhang \inst{1} \and Tianming Liu \inst{2}\and Li Su \inst{3,4}\and Dajiang Zhu\inst{1}\textsuperscript{(\Letter)}}
\authorrunning{ M. Chen et al.}
%
\institute{ Department of Computer Science and Engineering, University of Texas at Arlington, United States\\
\email{dajiang.zhu@uta.edu}\\
\and
School of Computing,  University of Georgia, United States\\
\and
Department of Psychiatry, University of Cambridge, United Kingdom\\
\and
Sheffield Institute for Translational Neuroscience, University of Sheffield, United Kingdom
}

\maketitle              
\begin{abstract}
Alzheimer's disease (AD) and Lewy body dementia (LBD) present overlapping clinical features yet require distinct diagnostic strategies. While neuroimaging-based brain network analysis is promising, atlas-based representations may obscure individualized anatomy.
Gyral folding-based networks using three-hinge gyri provide a biologically grounded alternative, but inter-individual variability in cortical folding results in inconsistent landmark correspondence and highly irregular network sizes, violating the fixed-topology and node-alignment assumptions of most existing graph learning methods, particularly in clinical datasets where pathological changes further amplify anatomical heterogeneity.
We therefore propose a probability-invariant random-walk-based framework that classifies individualized gyral folding networks without explicit node alignment. Cortical similarity networks are built from local morphometric features and represented by distributions of anonymized random walks, with an anatomy-aware encoding that preserves permutation invariance.
Experiments on a large clinical cohort of AD and LBD subjects show consistent improvements over existing gyral folding and atlas-based models, demonstrating robustness and potential for dementia diagnosis. 

\end{abstract}

\section{Introduction}
Alzheimer’s disease (AD) and Lewy body dementia (LBD) are two of the most prevalent neurodegenerative disorders, together accounting for a substantial proportion of dementia cases worldwide~\cite{nichols2022estimation}. Clinically, these two conditions share overlapping cognitive symptoms and neuropathological features, particularly in early stages, which often leads to diagnostic ambiguity. However, AD and LBD differ markedly in disease mechanisms, progression trajectories, and treatment responses, making accurate differential diagnosis crucial for patient management and therapeutic decision-making~\cite{matar2025biological,young2024data}. Developing neuroimaging-based biomarkers that can reliably distinguish AD from LBD therefore remains a central challenge in dementia research.
A growing body of evidence suggests that such neurodegenerative processes are not confined to isolated brain regions but instead manifest as system-level alterations involving distributed brain circuits. This view has motivated the modeling of the human brain as a complex, interconnected network, whose organization supports cognition and behavior~\cite{bressler2010large,park2013structural}. 
Within this network-based perspective, researchers represent the brain as a graph of interacting elements, and have extensively characterized dementia-related changes through alterations in structural and functional connectivity, revealing many network-level disruptions that cannot be fully explained by local atrophy alone~\cite{zhu2014connectome}.

Recently, gyral folding patterns have emerged as an alternative and biologically grounded basis for network construction~\cite{chen2026community,zhuang2025gyralnet}. In particular, the three-hinge gyrus (3HG), a reproducible folding configuration linked to cortical development and mechanical constraints, has been proposed as an individualized anatomical landmark for defining network nodes~\cite{chen2017gyral,he2022gyral,zhang2020cortical}. Gyral folding networks that use 3HGs as nodes and derive edges from structural connectivity (e.g., diffusion MRI tractography) or functional connectivity (e.g., resting-state fMRI synchronization) have demonstrated strong discriminative power in dementia diagnosis.
Several recent studies have highlighted the advantages of gyral folding-based networks over conventional atlas-based representations. Prior work~\cite{lyu2024mild} has shown that networks constructed from gyral folding landmarks can outperform atlas-based brain networks in early AD detection, suggesting that folding-derived nodes capture disease-sensitive structural patterns that are blurred by coarse parcellation. More recent efforts~\cite{chen2025unified,chen2025representing} have further integrated gyral folding networks with atlas-based representations in a hierarchical manner, using representation learning to construct a unified, continuous staging framework spanning AD and LBD. These studies collectively underscore the potential of folding-informed network models for dementia characterization.
Despite these advances, existing gyral folding network methods face several limitations. First, due to pronounced inter-individual variability in cortical folding, it is generally difficult to establish reliable one-to-one correspondences between folding landmarks across subjects~\cite{chen2025using,zhang2023cortex2vector,zhang2020identifying}. As a result, node identities are often only weakly aligned, rendering downstream graph learning models sensitive to arbitrary node indexing or residual spatial noise. Second, the number of detected folding landmarks may vary across individuals, leading to graphs with different sizes and topologies. These properties violate the fixed-node and strict-alignment assumptions implicit in most atlas-based and conventional graph learning approaches~\cite{kan2022brain,kawahara2017brainnetcnn}, limiting their applicability and robustness in clinical settings.

To address these challenges, we propose a Probability-Invariant Random-Walk Learning framework (PaIRWaL) for the classification of gyral folding networks. The method constructs cortical similarity networks from morphometric feature similarity within local neighborhoods of gyral folding landmarks, and models each network through distributions of anonymized random walks, thereby achieving graph isomorphism invariance in a probabilistic sense. Random walk sampling is guided by a minimum degree local rule (MDLR) to balance exploration across heterogeneous node degrees.
We further propose an Anatomy-Aware Anonymized Walk Recording module ($\text{A}^3$WR), in which each random walk is represented as an event sequence encoding structural transitions and neighborhood relations, while anatomical priors are incorporated via region-of-interest (ROI) attribute tokens under a permutation-invariant formulation. 
We validate the proposed framework on a clinical cohort of AD and LBD subjects, where it consistently outperforms existing gyral folding network baselines as well as atlas-based brain network classification models, demonstrating its effectiveness and robustness for dementia diagnosis.
\begin{figure}[htb]

  \centering
  \centerline{\includegraphics[width=\linewidth]{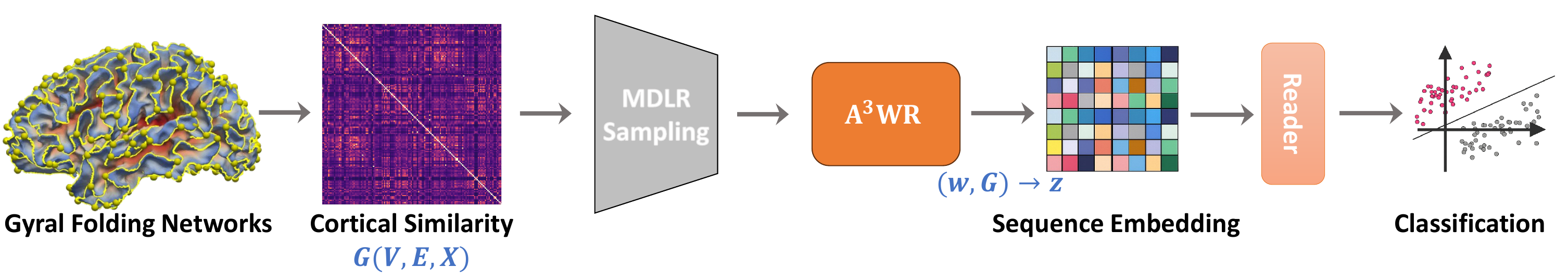}}
\caption{Overview of PaIRWaL. Random walks on gyral folding-based similarity networks are encoded into invariant sequences and aggregated for graph-level classification. }
\label{fig:overview}
\end{figure}


\section{Materials and Methods}
\subsubsection{Dataset description and data pre-processing}
In this study, we analysed a cohort of T1-weighted structural MRI scans (n = 303) obtained from the University of Cambridge. A subset of the data originates from the Multimodal Imaging in Lewy Body Disorders (MILOS) project. The cohort consisted of 108 cognitively normal controls (NC), 90 participants with a diagnosis of AD, and 105 with LBD.
After denoising and bias field correction of low-quality images, cortical surface reconstruction and parcellation were performed using the FreeSurfer software suite~\cite{FISCHL2012774}. 
Each subject’s preprocessed cortical surfaces were mapped onto a spherical coordinate system and subsequently resampled to the standard fsaverage7 template, corresponding to the seventh icosahedral subdivision with 163,842 vertices per hemisphere.
The 3HGs were then automatically extracted on the white matter surface using the fully automated pipeline described in~\cite{cao2026gyral,chen2017gyral}.
Following the approach proposed in~\cite{sebenius2023robust}, we constructed cortical similarity networks based on these folding landmarks. For each 3HG, cortical morphometric features, including cortical thickness, curvature, sulcal depth, surface area, and gray matter volume, were extracted within a 5-ring neighborhood. Inter-node similarity was then quantified using a symmetric Kullback-Leibler divergence with multivariate kernel density estimation resampling.
Based on the resulting similarity measure, we constructed an undirected and unweighted graph
$G=(V,E)$, where $V=\{v_1,\ldots,v_n\}$ denotes the set of detected 3HG landmarks.
To ensure graph connectivity while suppressing spurious connections, we first extracted the minimum spanning tree and then retained the top 10\% of the strongest remaining similarity edges to define $E$.
For node attributes, we adopted a distribution-based morphological fingerprinting strategy, similar in spirit to BrainPrint~\cite{wachinger2015brainprint}. Each 3HG was represented by an 81-dimensional feature vector.
\subsection{Preliminaries and method overview}
We study graph-level classification on individualized gyral folding networks derived from structural MRI.
Each subject is represented by a graph $G=(V,E,X)$, where nodes $V$ correspond to 3HGs, edges $E \subseteq V \times V$ encode cortical similarity, and
$X \in \mathbb{R}^{n \times d}$ contains node-wise morphometric fingerprints.
Due to substantial inter-subject variability in cortical folding patterns, graphs may differ in both node cardinality and ordering, and no canonical correspondence exists across subjects.
Consequently, the desired classifier must be invariant to node permutations and robust to variable graph sizes.
In our framework, graph representations are obtained via random walk sampling, rendering the classifier stochastic.
We therefore model graph classification using a randomized predictor $X_\theta : \mathcal{G} \rightarrow \mathbb{R}^C$, where $X_\theta(G)$ is a random variable.
We say that $X_\theta$ is \emph{probability invariant} if, for any two isomorphic graphs $G \simeq H$, $X_\theta(G) \;\overset{d}{=}\; X_\theta(H)$, where $\overset{d}{=}$ denotes equality in distribution.
This definition provides a principled formulation of permutation invariance for stochastic graph models.

As shown in Fig.~\ref{fig:overview}, our proposed PaIRWaL consists of three components:
(i) a conductance-based random walk sampling scheme,
(ii) an anatomy-aware anonymized walk recording function that maps walks to sequences, and
(iii) a reader neural network for graph-level prediction. 
These components are described in detail in the following sections.

\subsection{Conductance-based random walk sampling}
As discussed in~\cite{kim2025revisiting}, a random walk algorithm is said to be \emph{probability invariant} if,
for any pair of isomorphic graphs $G \simeq H$ with isomorphism $\pi:V(G)\rightarrow V(H)$,
\begin{equation}
(v_0,\ldots,v_\ell) \;\overset{d}{=}\; (\pi(v_0),\ldots,\pi(v_\ell))
\label{eq:rw_invariance}
\end{equation}
This condition ensures that the distribution of sampled walks depends only on graph structure and not on vertex identities.
We generate random walks $(v_0,\ldots,v_\ell)$ on $G$ as a first-order Markov chain with transition probabilities
\begin{equation}
\mathbb{P}(v_t = u \mid v_{t-1} = w)
= \frac{c_G(w,u)}{\sum_{x \in \mathcal{N}(w)} c_G(w,x)}
\label{eq:rw}
\end{equation}
where $c_G:E(G)\rightarrow\mathbb{R}_{+}$ is an edge conductance function.
In this work, we instantiate $c_G$ using the \emph{minimum degree local rule (MDLR)}~\cite{abdullah2016speeding,david2018random} strategy:
\begin{equation}
c_G(u,v) := \frac{1}{\min(\deg(u),\deg(v))}
\label{eq:mdlr}
\end{equation}
Since vertex degrees are preserved under graph isomorphism, MDLR satisfies
$c_G(u,v) = c_H(\pi(u),\pi(v)), \quad \forall G \simeq H$, thus ensuring that the induced random walk is probability invariant.
In practice, we further impose a non-backtracking constraint $v_{t+1}\neq v_{t-1}$ to promote exploration, without affecting invariance.

\subsection{Anatomy-aware anonymized walk recording}
Given a random walk $w=(v_0,\ldots,v_\ell)$ sampled on a graph $G$, we define a recording function
$q:(w,G)\mapsto z$, which maps the walk to a machine-readable event sequence.
The sequence $z$ is constructed incrementally along the walk and encodes anonymized structural
transitions, local neighborhood relations, and node-associated attributes.
As shown in~\cite{kim2025revisiting}, the recording function $q$ is invariant if, for any pair of
isomorphic graphs $G\simeq H$ with isomorphism $\pi$,
\begin{equation}
q(v_0\rightarrow\cdots\rightarrow v_\ell, G)
=
q(\pi(v_0)\rightarrow\cdots\rightarrow\pi(v_\ell), H)
\label{eq:recording_invariance}
\end{equation}
To satisfy this, vertex identities are anonymized via a walk-dependent mapping $\phi_w$, which assigns each vertex a unique identifier according to its order of first
appearance along the walk, in contrast to identity-based schemes such as DeepWalk~\cite{perozzi2014deepwalk} and node2vec~\cite{grover2016node2vec}.
To enrich structural expressiveness while preserving invariance, we additionally record
\emph{named neighbors} discovered during the walk.
Let $\mathcal{S}_t$ denote the set of vertices visited up to step $t$.
When visiting $v_t$, we encode incident relations between $v_t$ and previously discovered neighbors $u\in\mathcal{N}(v_t)\cap\mathcal{S}_t$ whose connections have not yet been recorded, thereby capturing local subgraph structure beyond simple sequential transitions.
We further extend the anonymized encoding with anatomical priors.
Let $a:V\rightarrow\mathcal{A}$ denote a mapping from vertices to ROI labels.
When a vertex $v$ is visited, its corresponding ROI attribute token $a(v)$ is attached.
Since $a(v)=a(\pi(v))$ for any graph isomorphism $\pi$, incorporating ROI tokens does not violate
Eq.~\eqref{eq:recording_invariance}.
In addition, each vertex $v$ is associated with a morphometric fingerprint
$x_v\in\mathbb{R}^d$, which is injected as a continuous embedding aligned to the walk position.
Overall, the recording function combines walk-induced anonymization, local adjacency relations,
and intrinsic node attributes into a sequential representation.
Let $\Vert$ denote sequence concatenation. The resulting event sequence is given by
$q(w,G)
=
\big\Vert_{t=0}^{\ell}
\Big[
\phi_w(v_t),\;
\mathcal{N}(v_t)\cap\mathcal{S}_t,\;
a(v_t),\;
x_{v_t}
\Big]$.

\subsection{Reader neural network and aggregation}
The encoded sequence $z$ is processed by a reader neural network $f_\theta : z \mapsto \hat{y} \in \mathbb{R}^C$, which is the only trainable component of the framework.
Importantly, $f_\theta$ can be instantiated by an arbitrary sequence model, as probability invariance of the overall method is guaranteed by the random walk sampling and recording stages, independently of the specific choice of $f_\theta$.
Graph-level predictions are obtained by Monte Carlo aggregation over $K$ random walks, $\hat{y}_G = \frac{1}{K} \sum_{k=1}^K f_\theta(z_k)$, yielding a permutation- and size-robust estimator.
In this work, we implement $f_\theta$ as a lightweight two-layer MLP.
Model parameters are optimized using the cross-entropy loss between predicted and ground-truth labels.
\section{Experiments and Results}
\subsection{Experimental setup}
\noindent\textbf{Study design.}
To evaluate the performance of the proposed PaIRWaL framework, we report results from 10-fold cross-validation on four classification tasks: AD/LBD/CN multiclass classification, AD vs. CN, LBD vs. CN, and AD vs. LBD binary classification. Performance is assessed using accuracy (ACC), sensitivity (SEN), specificity (SPE), area under the ROC curve (AUC), and Matthews correlation coefficient (MCC).

\noindent\textbf{Baselines.}
We compare the proposed PaIRWaL against 5 graph-based models constructed on gyral folding networks and 3 atlas-based graph learning approaches.
For gyral folding network-based baselines, we consider:
(1) a graph statistics–based classifier using a linear support vector machine (GraphStat);
(2) three widely used graph neural network (GNN) models: GraphSAGE~\cite{hamilton2017inductive}, GIN~\cite{xu2018powerful}, and GAT~\cite{veličković2018graph}; and
(3) a graph convolutional network with anatomy-aware mean pooling (AMP-GCN)~\cite{chen2025unified}.
For atlas-based graph learning methods, we construct cortical similarity networks using the Destrieux atlas as the parcellation scheme, and evaluate three representative approaches: BrainNetCNN~\cite{kawahara2017brainnetcnn}, BNT~\cite{kan2022brain}, and CPSSM~\cite{chen2025core}.

\noindent\textbf{Implementation details.}
During training, $K=8$ walks per graph were sampled, while 64 walks per graph were used at inference.
And the length of walks was set to $\ell=64$.
An MLP reader with a hidden dimension of 128 and a dropout rate of 0.2 was optimized using AdamW (learning rate $10^{-3}$, weight decay $10^{-4}$) for 200 epochs, and early stopping with a patience of 30 epochs. The batchsize was set to 64 and all experiments were conducted on a server equipped with an Intel Xeon Gold 6442Y CPU and 2 NVIDIA H100 GPUs.
\subsection{Results}
\begin{figure}[h!]

  \centering
  \centerline{\includegraphics[width=\linewidth]{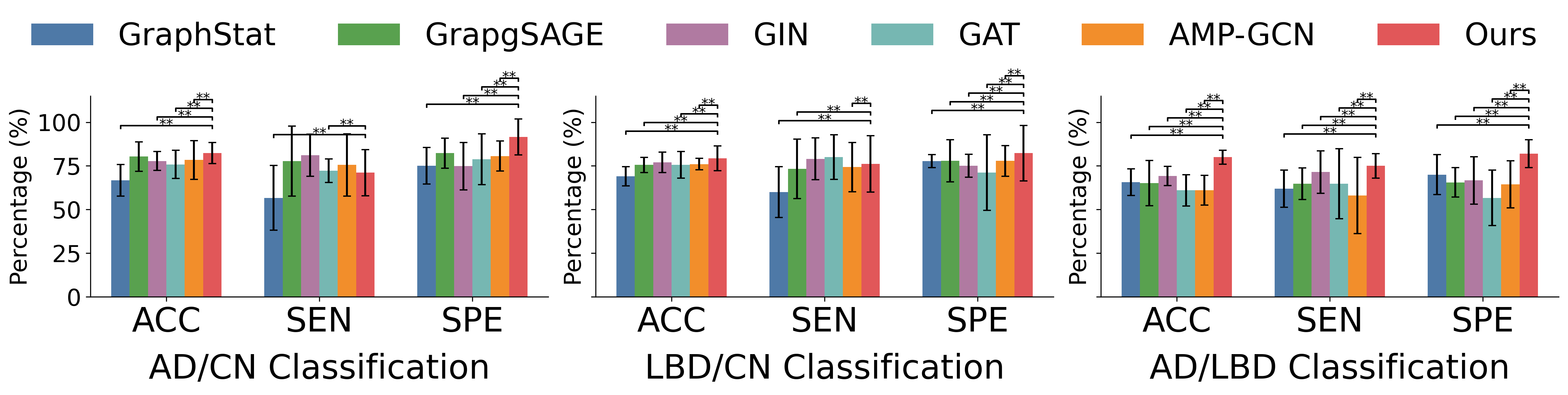}}
\caption{Classification performance for AD/CN, LBD/CN, and AD/LBD tasks.
Bars show mean ± standard deviation across cross-validation folds.
** denotes the statistical significance (p < 0.05) assessed using a two-sample t-test. }
\label{fig:binary}
\end{figure}

The performance comparison between the proposed PaIRWaL and the baselines is summarized in Table~\ref{tab:ablation}. PaIRWaL consistently achieves the best performance across all evaluation metrics, outperforming all competing approaches. Notably, it surpasses both folding network-based GNN models and atlas-based methods, demonstrating the effectiveness of probability-invariant random walk representations for handling individualized cortical folding variability.
Ablation studies further validate the contribution of each component. Removing backtracking, named neighbors, anonymization, or ROI attribute tokens consistently degrades performance, with the largest drop observed when anonymization or anatomical tokens are excluded. These results highlight the importance of invariant walk sampling, structure-aware encoding, and anatomical priors in achieving robust dementia classification.
The performance comparison on the three binary classification tasks (AD/CN, LBD/CN, and AD/LBD) is illustrated in Fig.~\ref{fig:binary}. 
\begin{table}[h!]
\centering
\caption{Performance comparison and ablation on AD/LBD/CN classification.}
\label{tab:ablation}
\setlength{\tabcolsep}{2.0pt}
\renewcommand{\arraystretch}{1.2}
\begin{tabular}{lccccc}
\hline
Method & ACC(\%) & SEN(\%) & SPE(\%) & AUC(\%)& MCC(\%) \\
\hline
BrainNetCNN & 64.02±7.17 & 63.26±6.82 & 81.88±3.58 & 83.88±5.44 & 46.17±10.77\\
BNT& 67.32±3.69& 66.00±3.22 & 83.44±1.85& 85.21±3.82 & 51.44±5.59\\
CPSSM & 69.64±2.36 & 68.61±2.61 & 84.65±1.27 & 84.30±3.35 & 55.03±3.68\\
\hline
GraphStat & 55.13±5.58 & 54.81±5.75 & 77.46±3.01 & 70.97±5.94 & 33.12±8.69\\
GraphSAGE& 62.40±4.33 & 61.56±4.59 & 81.09±2.11& 78.70±3.82 & 43.41±6.61\\
GIN & 61.43±8.39 & 60.86±8.25 & 80.66±4.05 & 79.06±5.67 & 42.51±12.40\\
GAT & 61.03±6.66 & 60.25±6.16 & 80.48±3.22 & 77.40±3.68 & 42.04±9.83\\
AMP-GCN & 60.72±2.86 & 60.46±2.45 & 80.40±1.36 & 78.73 ± 3.63 & 42.09±3.39\\
\hline
w/ backtracking& 66.34±3.74 & 65.41±3.34 & 82.98±1.84& 83.56±4.20 & 49.77±5.94\\
w/o NN & 67.01±6.68 & 65.80±6.27 & 83.29±3.32 & 84.38±4.02 & 50.88±10.57\\
w/o AN & 66.66±4.32 & 65.48±3.88 & 83.14±2.04& 83.83±4.78 & 50.60±7.03\\
w/o ROI& 69.31±2.99 & 68.30±2.73 & 84.49±1.45 & 85.51±3.11 & 54.17±4.62\\
\hline
PaIRWaL & \textbf{71.93±4.59} & \textbf{71.04±4.37} & \textbf{85.86±2.32} & \textbf{85.59±3.93} & \textbf{58.22±6.95}\\
\hline
\end{tabular}
\footnotesize{
NN: named neighbors; AN: anonymization; ROI: Anatomical ROI token.}
\end{table}
Across all settings, PaIRWaL consistently achieves superior performance over competing gyral folding–based graph learning methods. 
Notably, on the more challenging AD/LBD task, which requires differentiating two dementia subtypes with overlapping symptoms, PaIRWaL shows the largest gains, achieving clear improvements in both sensitivity and specificity. Statistical analysis (two-sample t-test, $p<0.05$) confirms that these improvements are significant compared with baseline models. 
These results further validate the effectiveness of the proposed probability-invariant random walk representation for reliable dementia diagnosis under diverse binary settings.

\begin{figure}[h!]

  \centering
  \centerline{\includegraphics[width=\linewidth]{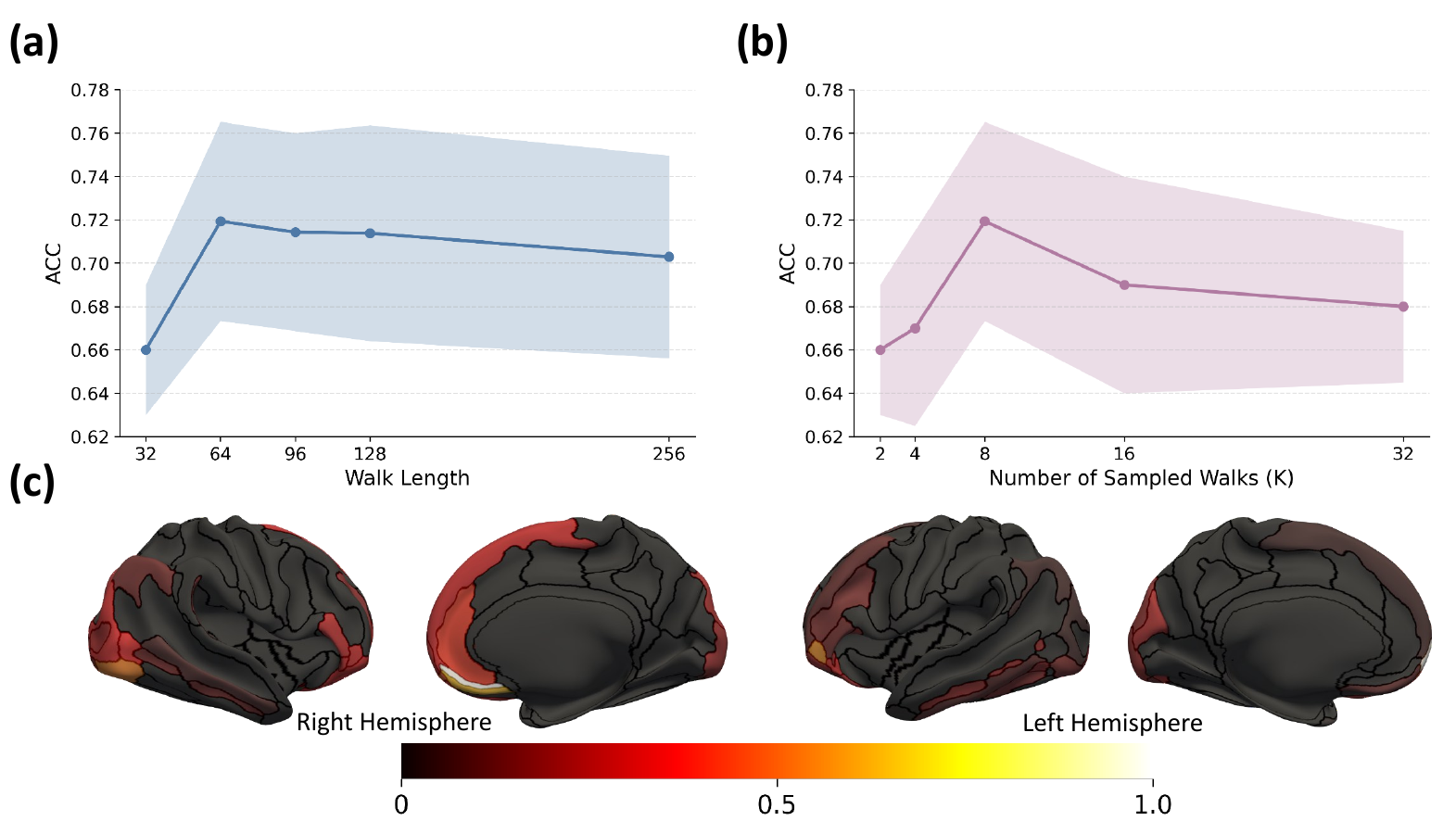}}
\caption{Sensitivity analysis and visualization of the proposed PaIRWaL.
(a) Classification accuracy under different random walk lengths.
(b) Performance with varying numbers of sampled walks per graph.
(c) Cortical surface projection of aggregated 3HG heatmaps, obtained by region-wise weighted averaging and normalization on the Destrieux atlas, highlighting discriminative folding regions. }
\label{fig:roi}
\end{figure}

\noindent\textbf{Sensitivity and interpretability studies.}
We further evaluate the sensitivity and interpretability of PaIRWaL, as illustrated in Fig.~\ref{fig:roi}.
As shown in Fig.~\ref{fig:roi}(a), classification performance improves as the random walk length $\ell$ increases and stabilizes around 64 to 128, suggesting that moderately long walks are sufficient to capture informative local and mesoscopic graph structures. 
Figure~\ref{fig:roi}(b) demonstrates a similar trend with respect to the number of sampled walks per graph, where performance peaks around 
$K=8$ and remains stable thereafter, indicating that only a small number of walks are required to obtain reliable graph-level estimates, thus ensuring computational efficiency.
Figure~\ref{fig:roi}(c) provides a cortical visualization of the aggregated 3HG heatmaps. 
The resulting projections reveal spatially localized and anatomically meaningful patterns concentrated in specific cortical folding regions. These findings suggest that the proposed model not only achieves strong predictive performance but also captures disease-relevant structural alterations, offering improved interpretability for dementia diagnosis.
\section{Conclusion}
We presented PaIRWaL, a probability-invariant random-walk learning framework for the classification of individualized gyral folding networks. This method models cortical similarity networks through anonymous random walk distributions rather than explicit node correspondences, naturally addressing individual differences in node identity and arbitrary network sizes. 
The conductance-based sampling and anatomy-aware encoding enable robust and permutation-invariant structural representations. Experiments on AD, LBD, and control cohorts demonstrate consistent improvements over existing approaches, highlighting the potential of the proposed framework for reliable dementia diagnosis.



%
%
%
\bibliographystyle{splncs04}
\bibliography{reference}
\end{document}